\documentstyle[11pt,aaspp,psfig]{article}
\def\gs{\mathrel{\raise0.35ex\hbox{$\scriptstyle >$}\kern-0.6em \lower0.40ex\hbox{{$\scriptstyle \sim$}}}}
\def\ls{\mathrel{\raise0.35ex\hbox{$\scriptstyle <$}\kern-0.6em \lower0.40ex\hbox{{$\scriptstyle \sim$}}}}
\def\et{\hbox{et al.}$\,$}

\begin{document}
\small
{\hfill \fbox{{\sc Revised version} \today}}

\title{The evolution of the galactic morphological types in clusters}
\author{
Giovanni Fasano,$^{\!}$\altaffilmark{1}
Bianca M.\ Poggianti,$^{\!}$\altaffilmark{1}
Warrick J.\ Couch,$^{\!}$\altaffilmark{2}
Daniela Bettoni,$^{\!}$\altaffilmark{1}
Per\ Kj\ae rgaard,$^{\!}$\altaffilmark{3}
Mariano\ Moles$^{\!}$\altaffilmark{4}
}
\smallskip

\affil{\scriptsize 1) Osservatorio Astronomico di Padova, vicolo dell'Osservatorio 5, 35122 Padova, Italy.}
\affil{\scriptsize 2) School of Physics, University of New South Wales, Sydney 2052, Australia.}
\affil{\scriptsize 3) Astronomical Observatory of Copenhagen University, Juliane Maries Vej 30, 2100 Copenhagen, DK}
\affil{\scriptsize 4) Instituto de Matematicas y Fisica Fundamental, CSIC, C/ Serrano 123, 28006 Madrid, Spain.}

\begin{abstract}

The morphological types of galaxies in nine clusters in the redshift
range $0.1\ls z\ls 0.25$ are derived from very good seeing
images taken at the NOT and the La~Silla--Danish telescopes, 
with all galaxies at $M_V < -20$ and within the central
$\sim$1 $\rm Mpc^2$ area being classified.
With the purpose of investigating the evolution of the 
fraction of different morphological types
with redshift, we compare our results with the morphological
content of nine distant clusters studied by the MORPHS group
(Dressler \et 1997), five clusters observed with
{\it HST-WFPC2} at redshift $z = 0.2-0.3$ (Couch \et 1998), and
Dressler's (1980) large sample of nearby clusters.
After having checked the reliability of our morphological
classification both in an absolute sense and relative to the MORPHS
scheme (Smail \et 1997), we analyze the relative occurrence of
elliptical, S0 and spiral galaxies as a function of the cluster
properties and redshift. We find a large intrinsic
scatter in the S0/E ratio, mostly related to the cluster morphology. 
In particular, in our cluster
sample, clusters with a high concentration of ellipticals display a
low S0/E ratio and, vice-versa, low concentration clusters have a
high S0/E. At the same time, the trend of the morphological fractions
(\%Es, \%S0s, \%Sp) and of the S0/E and S0/Sp ratios with redshift clearly
points to a morphological evolution: as the redshift decreases, the S0
population tends to grow at the expense of the spiral population,
whereas the frequency of Es remains almost constant.
We also analyze the morphology-density (MD) relation in our clusters and
find that -- similarly to higher redshift clusters -- a good MD
relation exists in the high-concentration clusters, while it is
absent in the less concentrated clusters. Finally, the comparison of
the MD relation in our clusters with that of the D97
sample suggests that the transformation of spirals into S0 galaxies
becomes more efficient with decreasing local density.

\end{abstract}


\sluginfo

\newpage\section{Introduction}

When Butcher \& Oemler (1978, 1984) discovered an excess of galaxies
bluer than the elliptical sequence in clusters at $z\ge0.2$, nothing
was known about the galactic morphologies at such large distances.
The first evidence of the disky/spiral nature of the Butcher-Oemler galaxies
came from high-resolution ground-based imagery which also found
several cases of galaxies with disturbed morphologies and/or close
neighbors (Thompson 1986, 1988, Lavery \& Henry 1988, 1994, Lavery,
Pierce \& McClure 1992).

Over the past five years, thanks to the high spatial resolution
imaging achieved with the {\it Hubble Space Telescope} {(\it HST)}, it
has been established that the morphological properties of
galaxies in rich clusters at intermediate redshift differ dramatically
from those in nearby clusters. The most obvious difference is the
overabundance of spirals in the cluster cores at $z=0.3-0.5$ (Couch et
al.\ 1994, 1998, Dressler et al.\ 1994, Wirth et al.\ 1994, Dressler
et al.\ 1997 [D97, MORPHS collaboration], Oemler et al.\ 1997, Smail
et al.\ 1997 [S97, MORPHS]). The spiral population in the distant
clusters consist of the great majority of the blue galaxies
responsible for the Butcher-Oemler effect, as well as a sizeable
fraction of the \sl red \rm population (Dressler et al.\ 1999 [MORPHS], 
Poggianti et al.\ 1999 [MORPHS]). A considerable
proportion of these spirals have disturbed morphologies, in some cases
quite clearly as the result of an ongoing merger/interaction
while in others possibly connected to some other dynamical mechanism
(e.g. interaction with the hot intracluster medium or the cluster tidal field,
Moore et al.\ 1996, 1998, Abadi et al. 1999).

The second major piece of evidence for morphological evolution in clusters 
was uncovered only from post-refurbishment data:
Coupled to the increase in the spiral fraction, the S0 galaxies
at intermediate redshifts are proportionately (x2--3) \sl less \rm
abundant
than in nearby clusters, while the fraction of ellipticals is already
as large or larger (D97, S97).  This result strongly suggests that a
large number of the cluster spirals observed at $z \sim 0.4$ have
evolved into the S0's that dominate the cores of rich clusters today
(D97, Couch et al.\ 1998, van Dokkum et al.\ 1998, but see Andreon
1998 for a different view).
Thus the disk galaxy populations appear to be greatly affected by
the cluster environment, while the ellipticals in dense regions seem to have
changed little since $z\sim 0.5$ as far as both their abundance and
their stellar populations are concerned (van Dokkum \& Franx 1996,
Andreon, Davoust \& Heim 1997, D97, Ellis et al.\ 1997 (MORPHS), Kelson et
al. 1997, S97, Barger et al.\ 1998 (MORPHS), van Dokkum et al.\ 1998,
Kelson et al.\ 1999).  
Morphological studies at redshift greater than 0.6 
have been limited to three
clusters so far, pointing to a low fraction of early-type galaxies in two
clusters at $z\sim 0.8$ (Lubin et al. 1998, van Dokkum et al. 2000), 
a high early-type galaxy fraction in a cluster at z=0.9 (Lubin et al. 1998)
and a surprisingly high
rate of mergers in a cluster at z=0.83 (van Dokkum et al. 1999).

Further proof of the changes occurring in clusters
is the observed evolution of the morphology-density (MD) relation -- 
the correlation between galaxy morphology and
local projected density of galaxies that Dressler (1980a, D80a) found 
{\it in all types of clusters} at low redshift, whereby the elliptical
fraction increases and the spiral fraction decreases with increasing
local galaxy density. An MD relation qualitatively similar to that found
by D80a was discovered by D97 to be present {\it in regular clusters and
absent in irregular ones at $z\sim 0.5$}.
Interestingly, the incidence of ellipticals is already very high
in all the distant clusters regardless of their dynamical status, therefore
the formation of the ellipticals must occur independently of and
before cluster virialization (D97).

Overall, the available data seem to require a strong
morphological evolution in clusters between $z=0.4$ and $z=0$. Still,
it is worth keeping in mind that these conclusions, although grounded
on high-quality data obtained with a monumental observational effort,
are based on a ``small'' 
sample of distant clusters and on the
comparison of a limited redshift range around $z\sim 0.4$ with the
present-day cluster populations (Dressler 1980b, D80b). 
Clearly the $z\sim 0.1-0.2$
regime - which up until now has remained largely 
unexplored - is crucial for a better
understanding of the progression of galaxy evolution in dense
environments. At these moderate redshifts, performing an analysis
comparable to that of the MORPHS requires either ground-based CCD imaging taken
over quite a large field under excellent seeing conditions, or a
time-consuming mosaic coverage with {\it HST}.

The goal of this paper is to begin to fill in the observational gap
between the distant clusters observed with {\it HST} and the
nearby clusters, and hence trace, for the first time, the evolution of
the morphological mix at a look-back time of $2-4$ Gyr. In addition, by
enlarging the sample of clusters whose galactic morphologies have been
studied in detail, we hope to shed some light on the dependence of the
observed evolutionary trends on the cluster properties. We present
ground-based, good-seeing images of the central regions of 9 clusters
at $z=0.09-0.25$ (\S2) and we perform a detailed morphological
analysis of the galaxies in these clusters (\S3).  We study the
relative occurrence of ellipticals, S0's and spirals as a function of
the cluster properties and we compare them with similar studies at
lower and higher redshift (\S4). Finally, we examine the morphology-density 
relation of the total sample and of the high- and low-concentration
clusters separately (\S5) and we present our conclusions in \S6. 
Throughout this paper we use $H_0=50
\rm \, km \, sec^{-1} \, Mpc^{-1}$ and $q_0=0.5$.

\section{Observations and sample selection}

\begin{table*}
{\scriptsize
\begin{center}
\centerline{\sc Table 1}
\vspace{0.1cm}
\begin{tabular}{ccccccc}
\hline\hline
\noalign{\smallskip}
 {Run} & {Date} & {Instrument} & {Pixelsize} & {Field} & {gain} & {r.o.n.} \cr 
       &        &             & (arcseconds)& (pixels)&        & \cr 
\hline
\noalign{\medskip}
  1  & 1995~May & STAN-CAM         & 0.176       & 1024$^2$& 1.69   & 6.36 \cr
  2  & 1995~Jun & STAN-CAM         & 0.176       & 1024$^2$& 1.69   & 6.36 \cr
  3  & 1997~Jan & DFOSC & 0.420    & 2052$^2$& 1.31   & 4.90 \cr
  4  & 1997~Feb & ALFOSC         & 0.187       & 2048$^2$& 1.02   & 5.60 \cr
\noalign{\smallskip}
\noalign{\hrule}
\noalign{\smallskip}
\end{tabular}
\end{center}
}
\vspace*{-0.8cm}
\end{table*}

The data presented here we taken as part of a long term project,
involving
four of us (GF,DB,PK,MM), aimed at analysing the scaling relations
of early--type galaxies in 25 clusters spanning the redshift range
$0.03-0.25$ (Fasano et al. 2000).  Only Abell clusters having Bautz
and Morgan (1970) types II or larger, Rood and Sastry (1971) types C or
F, and galactic latitude $\vert b\vert > 40^\circ$ were included in the
original sample.  Moreover, very poor clusters (Abell richness
class~=~0) were excluded from the selection.

The observations, taken in two or three bands ($B$, $V$, Gunn $r$), 
were collected at the NOT (STAN--CAM or ALFOSC) and 1.5~Danish (DFOSC)
telescopes
during four different observing runs from May~1995 to Feb~1997. 
A log of the observations pertinent to the clusters discussed
in this paper, together with the main properties of the CCDs
used, are presented in Table 1. 
The seeing at the NOT telescope ranged from 0.5 to 0.82
arcsec, except during run (4) when it was 1.1 arcsec; at the
1.5~Danish telescope (run 3) the seeing was $\sim 1.5$ arcsec.
However, the rest-frame resolution in kpc does not vary much within
the sample (see Table 2). Several standard star fields (Landolt 1992)
were observed during each night in the three bands in order to
set the proper photometric calibrations as a function of the zenith
distance and of the $B-G_r$ (or $V-G_r$) colors. Bias subtraction and
flat--fielding, together with the removal of bad columns and cosmic ray
events, were performed using the $CCDPROC$ tool in IRAF. A more detailed
description of the observations and data reduction procedures can be
found in Fasano et al.~(2000).

To be consistent with previous
morphological studies (D97), we should only consider clusters for
which at least the central 1~$\rm Mpc^2$ has been imaged. 
Due to the limited angular size of our CCD frames, we have excluded
from the present sample those clusters with $z < 0.09$, for which the coverage
turned out to be inadequate. Moreover,
since our observations were not conceived in order to satisfy the above
mentioned criterion, even for clusters with $z > 0.1$ the sampled area is sometimes
less than 1~$\rm Mpc^2$ and it often turns out to be shifted with respect
to the geometrical center of the cluster. 
Hence, among the 25 clusters observed as part of the scaling-relation 
program, we have selected 9 clusters in the redshift range $0.1\ls
z\ls 0.25$ for which an acceptable coverage of the central region has
been secured.

The basic information concerning the selected clusters and the
parameters relevant for our analysis are reported in Table 2. 
For two clusters, A2658 and A1878, the available frames cover about
half of the requested area: in the following we will explain how we
have tried to account for the partial coverage. For
Abell~2192 we list two different entries in Table 2: the first one
(run 1) refers to four contiguous Gunn~$r$ images covering a quite large
area, the second one (run 2) relates to a smaller part of the
cluster, for which ($B$-$r$) colors of galaxies are available, thus
allowing a more accurate photometric calibration.  When comparing
with the other clusters, we will refer to the large area image of
A2192, while we will make use of the smaller area to assess how strongly the
different coverage can affect the results.

\begin{table*}
{\scriptsize
\begin{center}
\centerline{\sc Table 2}
\vspace{0.1cm}
\begin{tabular}{lccccccccc}
\hline\hline
\noalign{\smallskip}
 {Cluster} & z & RA         & DEC        & E(B-V) & Run & area    & seeing & seeing & $M_{lim}$ \cr 
           &   & (J2000)    &    (J2000) &           &        & $\rm Mpc^2$ &  arcsec   &  kpc   & (Gunn r) \cr 
\hline
\noalign{\medskip}
 A3330   & 0.091 &05$^h$~14$^m$~47$^s$&-49$^{\circ}$~04$^{\prime}$~19$^{\prime\prime}$&0.00& 3 &  1.0 & 1.53 & 3.46 & 18.36 \cr
  A389   & 0.116 &02$^h$~51$^m$~31$^s$&-24$^{\circ}$~56$^{\prime}$~05$^{\prime\prime}$&0.00& 3 &  1.2 & 1.45 & 4.04 & 18.85 \cr
  A951   & 0.143 &10$^h$~13$^m$~55$^s$ &+34$^{\circ}$~43$^{\prime}$~06$^{\prime\prime}$&0.01& 4 &  1.0 & 1.10 & 3.65 & 19.34 \cr
 A2658   & 0.185 &23$^h$~44$^m$~59$^s$ &-12$^{\circ}$~18$^{\prime}$~20$^{\prime\prime}$&0.10& 2 &  0.4 & 0.70 & 2.78 & 20.46 \cr
 A2192(l)& 0.187 &16$^h$~26$^m$~37$^s$ &+42$^{\circ}$~40$^{\prime}$~20$^{\prime\prime}$&0.01& 1 &  1.8 & 0.55  & 2.22 & 20.21 \cr
 A2192(s)& 0.187 &16$^h$~26$^m$~37$^s$ &+42$^{\circ}$~40$^{\prime}$~20$^{\prime\prime}$&0.01& 2 &  0.4 & 0.82  & 3.31 & 20.21 \cr
 A1643   & 0.198 &12$^h$~55$^m$~54$^s$ &+44$^{\circ}$~04$^{\prime}$~46$^{\prime\prime}$&0.00& 1 &  0.9 & 0.50 & 2.10 & 20.33 \cr
 A2111   & 0.229 &15$^h$~39$^m$~38$^s$ &+34$^{\circ}$~24$^{\prime}$~21$^{\prime\prime}$&0.06& 1 &  0.9 & 0.70 & 3.22 & 20.80 \cr
 A1952   & 0.248 &14$^h$41$^m$~04$^s$ &+28$^{\circ}$~38$^{\prime}$~12$^{\prime\prime}$&0.00& 2 &  1.2 & 0.60 & 2.93 & 20.92 \cr
 A1878   & 0.254 &14$^h$~12$^m$~49$^s$ &+29$^{\circ}$~12$^{\prime}$~59$^{\prime\prime}$&0.00& 1 &  0.6 & 0.50 & 2.47 & 20.92 \cr
\noalign{\smallskip}
\noalign{\hrule}
\noalign{\smallskip}
\end{tabular}
\end{center}
}
\vspace*{-0.8cm}
\end{table*}

\section{Galaxy catalogs and morphological classification}

Catalogs of galaxies for each frame have been obtained using
SExtractor (Bertin and Arnouts 1996). The galaxy magnitudes have
been corrected for a well known bias affecting SExtractor magnitudes
of galaxies having an $r^{1/4}$ profile (Franceschini et al. 1998).

As in D97, the analysis of the morphological types has been
done for galaxies down to a visual absolute magnitude $M_V \sim -20.0$.
The corresponding $r$-band magnitude limits were derived using standard
Gunn-$r$ and Cousins/Johnson ($V$) filter transmission, first
adopting the conversion between Gunn-$r$ and Cousins $R$ given by
Jorgensen (1994) and then applying  
the K-corrections of an intermediate energy distribution (Sab)
for $H_0=50 \rm \, km \, sec^{-1} \, Mpc^{-1}$ and $q_0=0.5$.
Examples of K-corrections for various galactic types can be found
in Poggianti (1997).
The magnitude limits were corrected for the foreground Galactic extinction 
(see Table 2) according to the standard Galactic extinction law (Mathis 1990).

Column 8 of Table 2 reports the seeing in arcseconds, whereas column 9
shows the rest-frame resolution that ranges between 2 and 4 kpc. Even
though the seeing quality of our imagery was often excellent, the
spatial resolution is poorer than that secured by the {\it HST}
imaging of the clusters in D97 ($\sim 0.7$~kpc at $z\sim$ 0.5), making
the merely visual classification less reliable with respect to that
given by the MORPHS. In order to improve the morphological type
estimates we have produced luminosity and geometrical (ellipticity and
position angle) profiles of all selected galaxies using the automatic
surface photometry tool $GASPHOT$ (Pignatelli and Fasano 1999). In this
way, in addition to the appearance of the galaxies on the images and
to the surface and isophotal plots, we took advantage of the typical
indications coming from the morphological profiles. For instance, S0
galaxies, even if poorly resolved, are usually characterized by
increasing ellipticity profiles (indicating an extended disk superimposed
upon an inner, round bulge), composite luminosity profiles (with
$r^{1/4}$ inner part and exponential outer part) and almost constant
position angle profiles in the outer part (disk). In contrast,
constant (or even decreasing) ellipticity profiles and outer isophotal
twisting (together with `pure' $r^{1/4}$ luminosity profile) are
highly suggestive of an elliptical morphology. An exponential
luminosity profile, with almost constant (or fluctuating) position
angle and ellipticity profiles are good hints of spiral
morphology. A break-down in ellipticity, coupled with a sharp (and
large) change of position angle, usually indicates the presence of a
bar, thus suggesting a spiral (or S0) classification. Clearly, the
above indications cannot be considered as unfailing rules, but
certainly they contributed to make the classifications more robust.
Table~3 (available on CD-ROM form) reports the positions and the
morphological classifications of all galaxies in our sample.

The morphological classification of the selected galaxies was
done by GF on the $r$-band images relying on both the visual appearance
and the profiles. 
Galaxies whose broad classification (E/S0/Sp) was judged uncertain
have been recorded with a question mark (eg: E?, S0?, Sp?) in Table~3.
The transition objects
(E/S0, S0/E, S0/a, Sa/0) have been `arbitrarily' assigned to some
broad class (E/S0/Sp) on the basis of the experience and of the
opinion of the classifier (GF). However, the relative number of
galaxies with uncertain and/or transition morphology turns out to add
a negligible contribution to the errorbars of the morphological
frequencies which are dominated by the Poissonian
uncertainties.

It is obviously crucial for our purposes to assess the reliability of
our classification scheme both in an absolute sense and relative to
the MORPHS scheme. To this end we have devised four different 
`blind' tests. The absolute accuracy of our classification has
been checked in two ways:

a) Using the $MKOBJECTS$ tool in IRAF, we have produced a set of `toy'
galaxies with different bulge/disk luminosity and size ratios and
varying inclination,
trying to reproduce E, S0, Sa and Sbc galaxies according to the typical
ratios given by Simien and de~Vaucouleurs (1986). 
The proper values of noise, seeing, pixel size and redshift have been 
used in the simulations to mimic, at best, the observing conditions of  
the cluster A2111 at z=0.23 which represents an average case for
its seeing and rest-frame resolution (0.7 arcsec, 3.22 kpc). The toy
galaxies have been then classified using the same tools and the same
classification scheme used for real galaxies.

b) The nearby galaxy imaging collection of Frei et al. (1996) has been
used to produce redshifted versions (with the $\sim (1+z)^4$ 
surface brightness dimming taken into account)  of 34 galaxies of
different morphological
types, including all the galaxies listed by Frei et al. to have 
types E to Sa (T=-5 to 1) and eight later-type galaxies (T=2 to 6).
In this way we intended to verify the capacity to discriminate
between ellipticals, S0s and early-type spirals and to recognize
later-type galaxies, the latter being an easier task than the former. 
Again the proper values of the observing parameters have been
used to mimic our A2111 images and the resulting galaxies have been
classified following the procedure and the rules used for our cluster
galaxies. 

\newpage

\hbox{~}
\centerline{\psfig{file=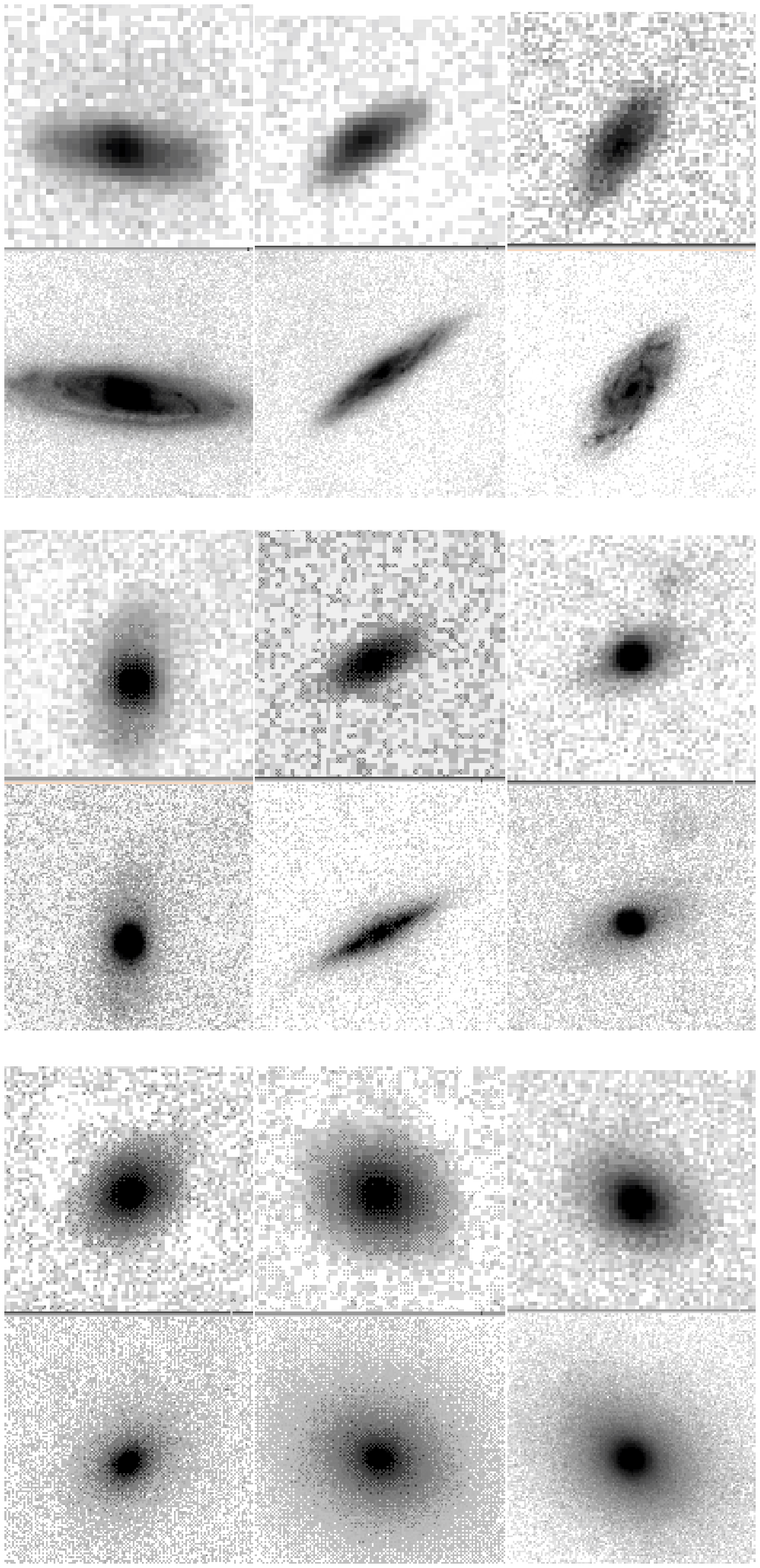,width=8cm}}
\vspace{-2.cm}
\noindent{\scriptsize
\addtolength{\baselineskip}{-3pt} 
\vspace*{3cm}\hspace*{0.3cm}

{\bf Fig.~1.(to be viewed in landscape)} -- 
\ Examples of Frei's nearby galaxy images (lower row of each set)
compared with the corresponding redshifted versions (upper row of 
each set). Three
galaxies for each broad morphological type
(E:S0:Sp) are presented. From bottom to top: the elliptical galaxies
NGC~4365, NGC~4472 and NGC~4636; the S0 galaxies NGC~3166, NGC~4710 and
NGC~4754; the spiral galaxies NGC~3623, NGC~3877 and NGC~6118.
\addtolength{\baselineskip}{3pt}
}

\newpage

In Figure~1 three examples of Frei's nearby galaxy
images for each broad class (E:S0:Sp) are shown, together with the
corresponding redshifted versions.

The main goal of these tests is to determine whether the quality
of our images is sufficient to recognize, at the cluster redshifts, the
salient features of the galactic morphologies, enabling us to broadly
classify galaxies into ellipticals (E,E?,E/S0), S0s (S0,S0?,S0/E,S0/a)
and spirals (Sa/0 and later) {\it as we would do for nearby galaxies}.
These tests show that in the great majority of cases (89\% in test [a]
and 73\% in test [b]) the broad
morphological types (E, S0 and spirals) assigned by GF at z=0 are also
recovered at $z\sim 0.23$ with no systematic shift among the types.
The only exception is represented by those galaxies from Frei's
catalog that were classified as S0/a at z=0. In fact, half of them
enter the ``spiral class'' as early spirals when viewed at z=0.23,
while in our classification the S0/a galaxies belongs to the S0 class.
Excluding these S0/a galaxies, low-z ellipticals, S0s and spirals 
were recovered at z=0.23 in the 92\% (11/12), 71\% (5/7) and 100\% (8/8)
of the cases, respectively.
Test a) shows that, regardless of the redshift, the seeing conditions, 
etc, S0 galaxies viewed face-on are classified ellipticals, while
the inclination of S0 and Sa galaxies influences the assignment to one
class or the other as edge-on S0s are easily mistaken for Sa's, while
Sa's at small inclinations are classified S0's. 

\hbox{~}
\centerline{\psfig{file=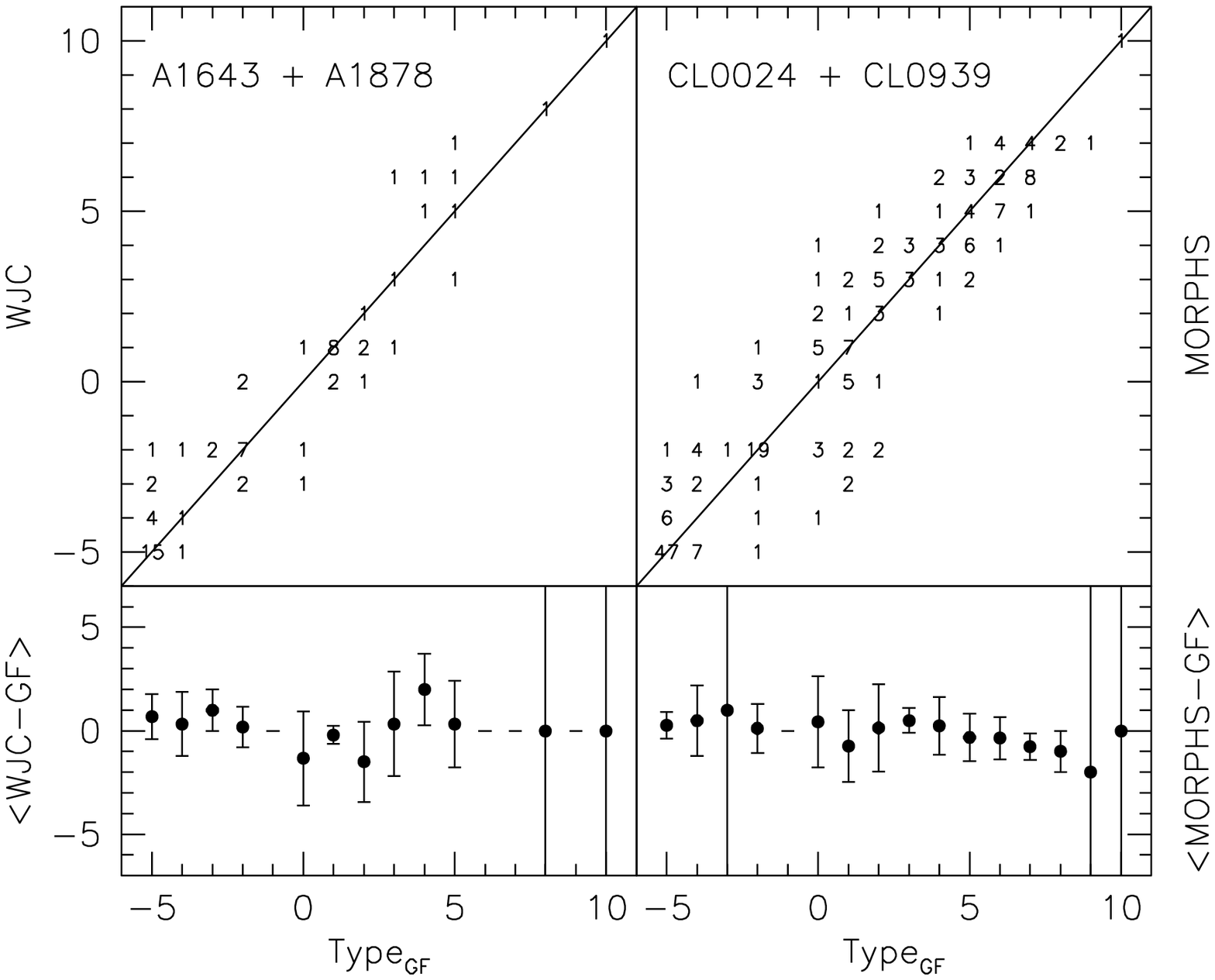,angle=0,width=5.0in}}
\vspace{-0.8cm}
\noindent{\scriptsize
\addtolength{\baselineskip}{-3pt} 
\vspace*{-2.5cm}\hspace*{0.3cm} 

{\bf Fig.~2.} -- 
\ {\sl Upper-left panel}: comparison between the de~Vaucouleurs T types
assigned by GF and WJC in their classifications of 
67 galaxies from the groundbased (NOT) imaging
of the clusters Abell~1643 and Abell~1878. The number of galaxies in
each bin of the GF versus WJC plane is indicated. {\sl Upper-right
panel}: 
comparison between the de~Vaucouleurs T types assigned by GF and the 
MORPHS in their $HST$--based classifications of the clusters CL~0024+16
and CL~0939+47.
{\sl Lower panels}: average differences between the external and the GF
types versus the GF classifications. Error bars illustrate the
statistical variances.
\addtolength{\baselineskip}{3pt}
}

In the tests described above, our types are compared with morphological
classifications which are considered correct {\it a priori}, since they
refer
to toy or nearby galaxies. However, in order to be able to compare our results
with higher redshift clusters, besides an absolute
check, we need to test the consistency between our classifications and
those of the MORPHS group. This has been done in two ways:

c) GF undertook independent visual classifications of galaxies in
the MORPHS {\it HST} images of the clusters CL~0024+16 and CL~0939+47.

d) WJC (one of the classifiers of the MORPHS collaboration) has
provided independent visual classifications of 67 galaxies in two
clusters of the present sample (Abell~1643 and Abell~1878).

The results of tests c) and d) are summarized in Figure~2.  At first
sight the agreement between GF and the external classifiers
seems to be quite good. However, after counting the total number of
galaxies which, according to the different classifiers, fall in the
different broad morphological types (E/S0/Sp), it was found that GF
had classified a smaller number of galaxies as S0 in comparison to
WJC and the MORPHS (57 versus 73).
Twelve of these galaxies GF had classified as ellipticals, while
the remaining 4 had been classified as spirals. 

We stress that these differences are actually not statistically
significant as far as the counts are concerned. In the worst case --
the S0 galaxies -- the Poissonian uncertainties are such that the
difference in numbers between GF and the other classifiers (57 cf. 73)
represents only a 1.5$\sigma$ marging. Nevertheless, to be conservative, 
we have assumed the above difference in assignment between the 
classifiers to be
systematic. In section 5, where an overall discussion of all
available data will be presented, in order to consistently compare our
data with those from the MORPHS sample, we will introduce a
statistical correction to account for these differences in the
morphological classification.

An indirect, fully independent confirmation of the correctness of
our morphological classifications may be obtained by looking at the
ellipticity distribution of elliptical and S0 galaxies in our
clusters. These distributions are shown in Figure~3, together
with the the ellipticity distribution of E galaxies derived by Fasano
and Vio (1991) for a large sample of local objects (smooth solid
line). The agreement with the corresponding distribution from our
elliptical galaxy sample is fairly good, supporting the assumption that we
are sampling the same population. Moreover, our E and S0 ellipticity
distributions are very similar to the corresponding ones shown in S97
and D97 for the MORPHS dataset, the Coma cluster (Andreon et al. 1996)
and the nearby cluster sample of D80b. It is worth noting that the
lack of round objects in the ellipticity distribution of S0 galaxies
confirms the previously mentioned bias in the classification of 
face-on S0s (see test [a]). This kind of misclassification,
however, has been shown to be almost unavoidable, even for nearby
galaxies (Capaccioli et al. 1991).

\hbox{~}
\centerline{\psfig{file=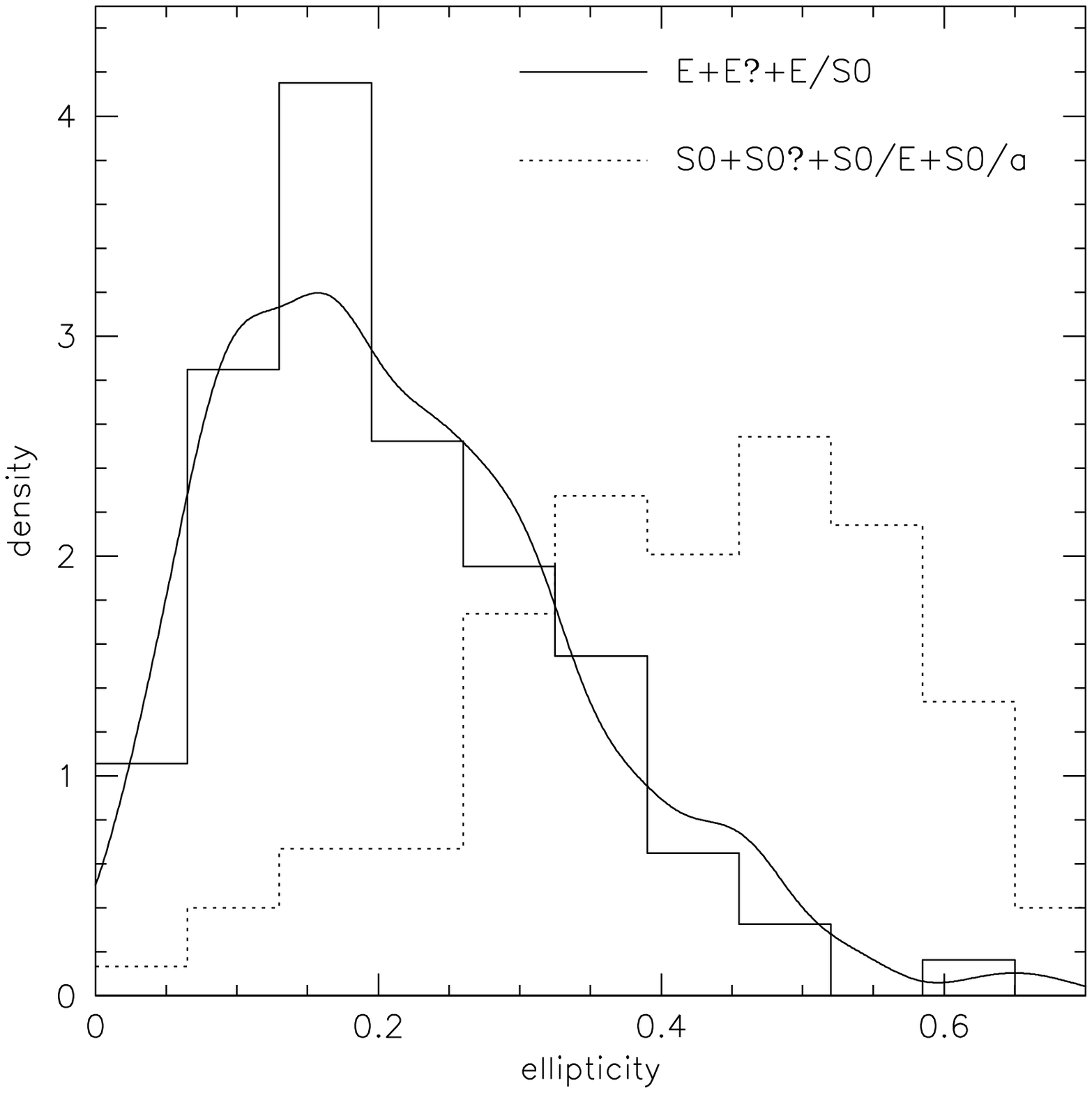,angle=0,width=3.7in}}
\vspace{-0.45in}
\noindent{\scriptsize
\addtolength{\baselineskip}{-3pt} 
\hspace*{0.3cm} {\bf Fig.~3.} \ Ellipticity distributions of elliptical
and S0 galaxies in our clusters. The smooth solid line represents the 
ellipticity distribution of Es derived by Fasano and Vio (1991) for a 
large sample of local objects. 
\addtolength{\baselineskip}{3pt}
}

\section{Results and comparison with other samples}

\begin{table*}
{\scriptsize
\begin{center}
\centerline{\sc Table 4: morphological counts of our clusters}
\vspace{0.1cm}
\begin{tabular}{lccccc}
\hline\hline
\noalign{\smallskip}
 {Cluster} & z & E & S0 & Sp & S0/E \cr 
\hline
\noalign{\medskip}
1) A3330 & 0.091 & 18 & 18 & 11 & 1.00$\pm$0.33 \cr
2) A 389 & 0.116 & 17 & 18 & 3  & 1.06$\pm$0.36 \cr
3) A951  & 0.143 & 10 & 12 & 2  & 1.20$\pm$0.50 \cr
4) A2658 & 0.185 & 11 &  6 & 3  & 0.54$\pm$0.29 \cr
5) A2192(l) & 0.187 & 17 & 18 & 14 & 1.06$\pm$0.36 \cr
5) A2192(s) & 0.187 &  9 & 12 &  5 & 1.33$\pm$0.59 \cr
6) A1643 & 0.198 & 15 & 17 & 15 & 1.13$\pm$0.40 \cr
7) A2111 & 0.229 & 34 & 17 & 16 & 0.50$\pm$0.14 \cr
8) A1952 & 0.248 & 25 & 14 & 14 & 0.56$\pm$0.20 \cr
9) A1878 & 0.254 & 15 & 8 & 15 & 0.53$\pm$0.23  \cr 
\noalign{\smallskip}
\noalign{\hrule}
\noalign{\smallskip}
\end{tabular}
\end{center}
}
\vspace*{-0.8cm}
\end{table*}

In Table~4 we list, for each cluster, the observed numbers of Es, S0s and
spirals and the observed S0/E ratio, together with the Poissonian
error. The clusters appear to be grouped in two different families,
according to their S0/E ratios: a low S0/E family (4 clusters) with
S0/E$\sim$0.5, and a high S0/E family (5 clusters) with S0/E$\gs$1.1.
We note that if we apply to our counts a statistical correction to
compensate for the excess of Es and the lack of S0s we have found in our
classifications with respect to the MORPHS (see \S 3), the S0/E
ratios of the above mentioned families both shift upwards, but the
dichotomy remains (see Figure~10a).
We have investigated whether this S0/E dichotomy could be driven by the small
cluster-to-cluster variations in the rest-frame characteristics of the
images, but the observed S0/E ratio is found to be uncorrelated with
the resolution and with the area surveyed as shown in Fig.~4a,b.
\footnote{The sensitivity of the S0/E ratio to the rest-frame area 
surveyed, was tested using A2192, which has large coverage. This
showed only a modest effect (see also Table 2).}
Hence, the S0/E dichotomy seems to reflect an intrinsic difference in 
the relative proportions of these types of galaxies in the two subsets 
of clusters.

\hbox{~}
\vspace{-2.5in}
\centerline{\psfig{file=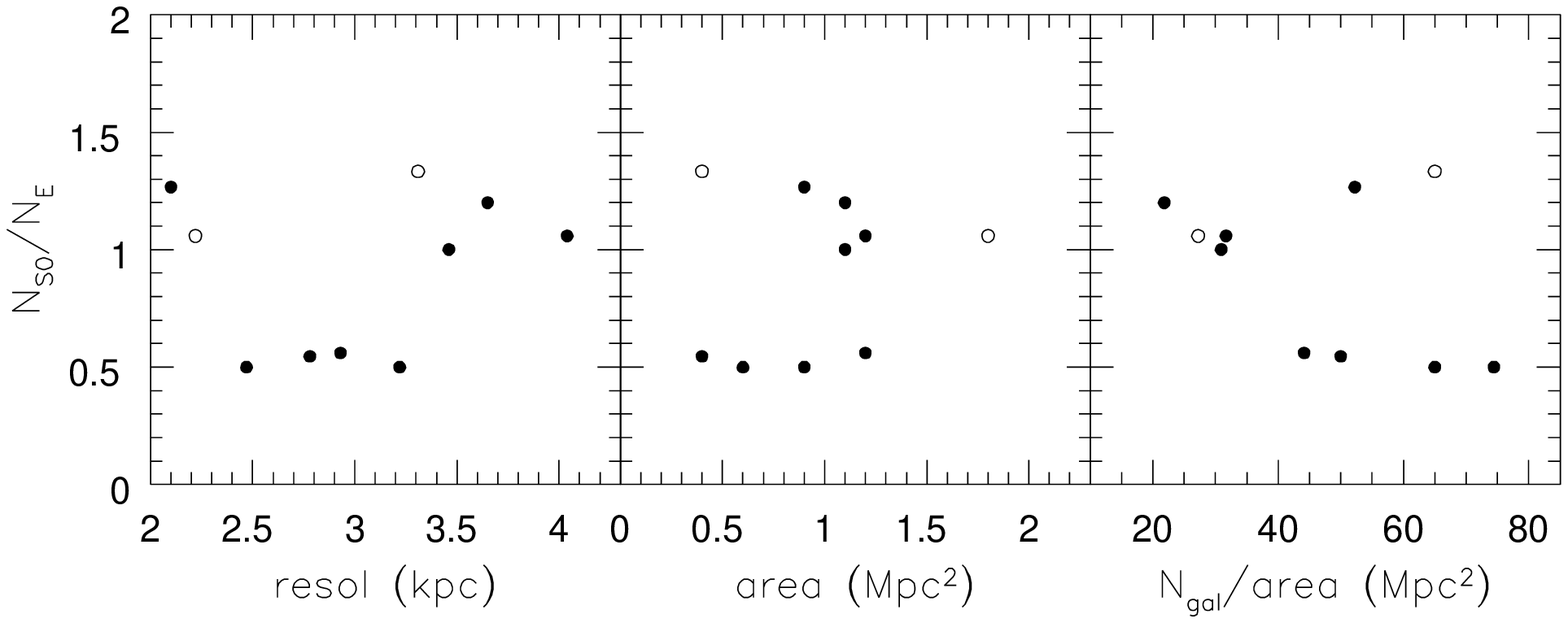,angle=0,width=5in}}
\noindent{\scriptsize
\addtolength{\baselineskip}{-3pt} 
\hspace*{0.3cm} {\bf Fig.~4.} \ 
The S0/E ratio versus the rest-frame resolution (in kpc), the area (in 
$\rm Mpc^2$) and the mean galaxy density (in numbers of galaxies per $\rm
Mpc^2$). The {\it open} circles refer to A2192, where both the smallest
area (0.4 $\rm Mpc^2$) and the largest area (1.8 $\rm Mpc^2$) values
are plotted.  
Errorbars are not displayed for the sake of clarity and can
be found in Table~4.
\addtolength{\baselineskip}{3pt}
}

The low S0/E and the high S0/E clusters are at $z \geq 0.19$ and $z 
\leq 0.2$, respectively,
but this step-like behaviour of the S0/E ratio at $z\sim 0.2$ is unlikely
to be an abrupt evolutionary effect: rather, it could be related to
different characteristics of the two families of clusters. Searching
for correlations between the S0/E ratio and the global cluster
properties, we have found no relation with the mean projected galaxy
density (Fig.~4c). {\it The only structural difference between the low-S0/E
and the high-S0/E clusters seems to be the presence/absence of a
high concentration of elliptical galaxies in a region that is
identified as the cluster centre.} This effect, visible even for
individual clusters, is evident in Figure~5, where the centered
\footnote{In the following the cluster centers are defined, for each 
cluster, by the median coordinates of all galaxies. However, the results 
turn out to be very similar if a different definition of the center is 
adopted (i.e mean rather than median coordinates and/or elliptical rather 
than whole population).}
maps of the low-S0/E and high-S0/E clusters are superimposed
separately on a rest--frame absolute scale. 

\hbox{~}
\vspace{-1in}
\centerline{\psfig{file=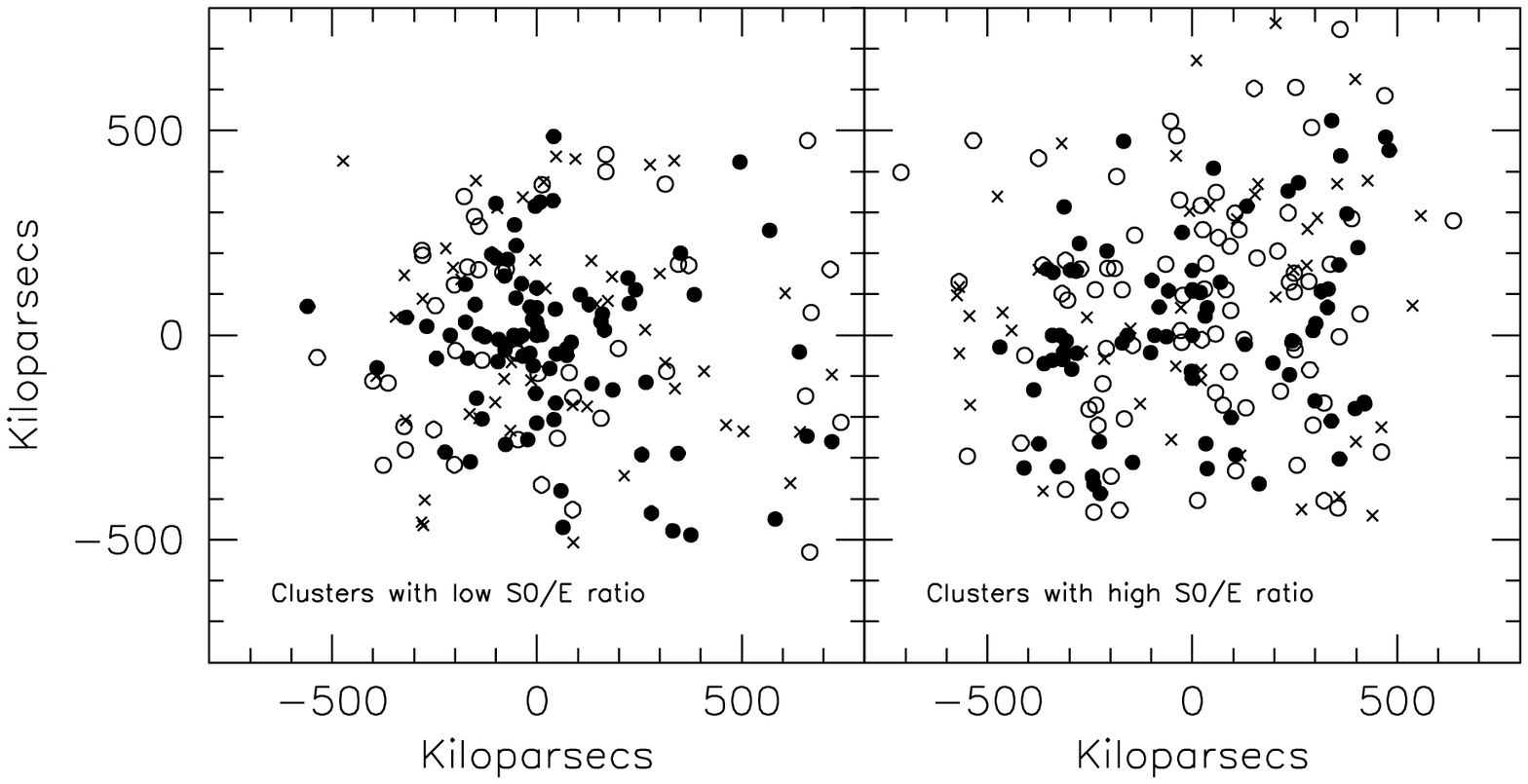,angle=0,width=5.2in}}
\vspace{-1in}
\noindent{\scriptsize
\addtolength{\baselineskip}{-3pt} 
\hspace*{0.3cm} {\bf Fig.~5.} \ 
Overlapping of the centered maps of the
low-S0/E and high-S0/E clusters. Filled dots, open dots and crosses
refer to ellipticals, S0s and spirals, respectively.
\addtolength{\baselineskip}{3pt}
}

A more quantitative illustration of the difference in the galaxy
spatial distribution between the low-S0/E and the high-S0/E clusters
is given in Fig.~6 using the Kolmogorov--Smirnov (KS) test. The cumulative
radial distribution of the ellipticals in the low-S0/E clusters is
significantly steeper (i.e. more concentrated) than that of the high
S0/E clusters. The two S0 distributions are practically
indistinguishable and those of the spirals do not differ
significantly, while the two total distributions of all types of galaxies
are dominated by the ellipticals and are statistically different.

\hbox{~}
\centerline{\psfig{file=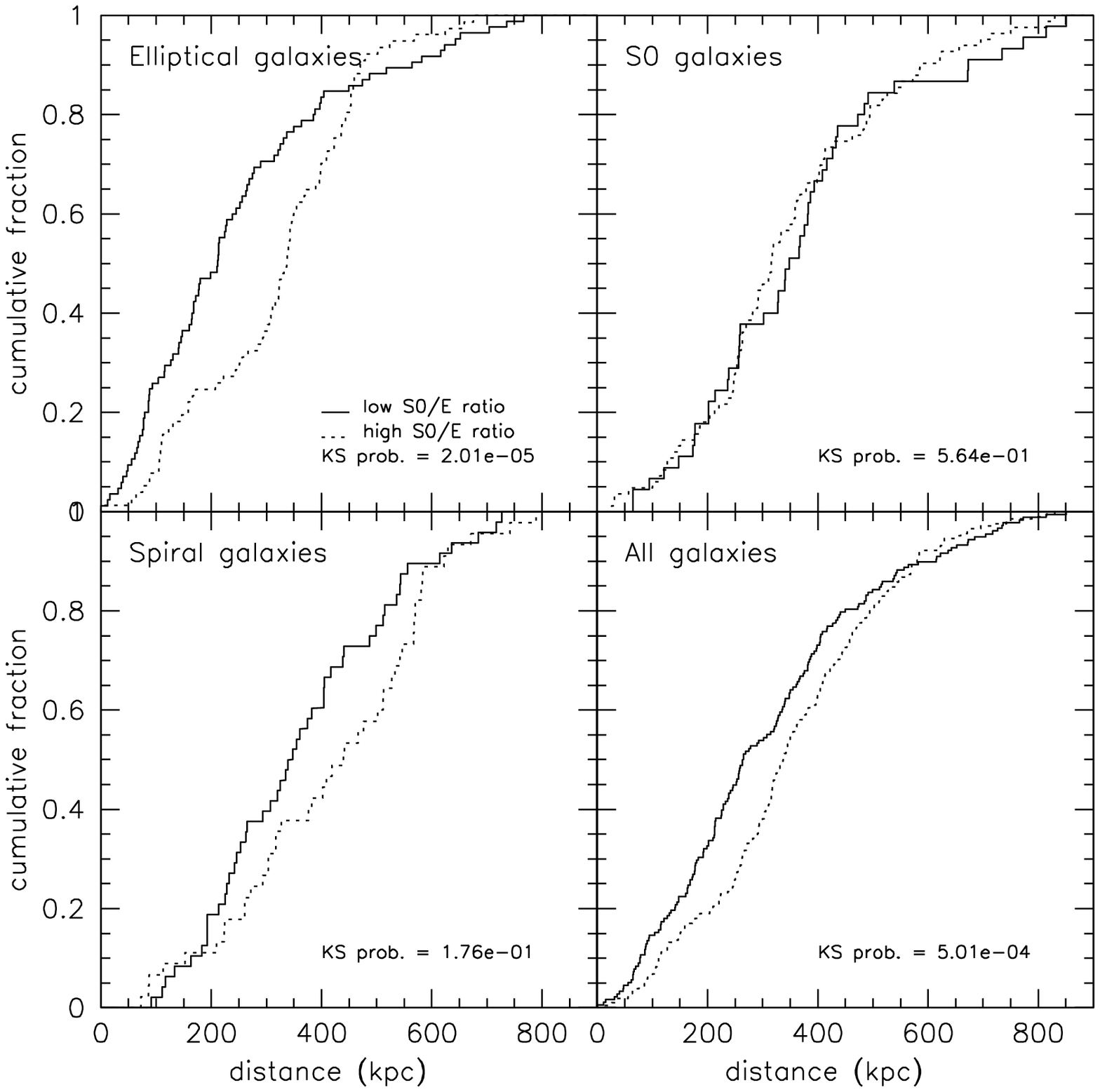,angle=0,width=5in}}
\noindent{\scriptsize
\addtolength{\baselineskip}{-3pt} 
\hspace*{0.3cm} {\bf Fig.~6.} \ 
Kolmogorov--Smirnov test applied to the overlapped radial
distributions (see Figure~5) of galaxies of different types for
low-S0/E (full lines) and high-S0/E (dotted lines) clusters.
\addtolength{\baselineskip}{3pt}
}

It is worth stressing that, both the `first sight' impression from
Figure~5 and the `objective' test in Figure~6 might actually be biased
by the differences in the rest--frame sampled areas among the
clusters, as well as by the slight off--centering of the CCD frames
with respect to the cluster centers and by the irregular shape of the
area surveyed. In an attempt to overcome this problem, we present in
Figure~7
a different representation of the cumulative radial distributions. In
this figure each line represents a different cluster (the numbers at
the end of each line identify the clusters according to the ranking in
Table~4) and the distributions are not normalized to the total number
of galaxies. Moreover, in order to correct for incompleteness due to
the irregular shape of the area surveyed, the cumulative numbers
$N_{CC}$ in Figure~7 are obtained by adding, for each new galaxy at
increasing distance $r$, the quantity $1/C(r)$, where $C(r)$ is a
completeness factor expressing the fraction of the circular area $\pi
r^2$ which is actually included in the area surveyed. These
corrections turn out to be small and never exceed $10\%$. Even
though not quantifiable by statistical tests like a KS, in Figure~7
the difference in the concentration of the elliptical population
between low-S0/E (full lines) and high-S0/E (dashed lines) clusters
stands out, with the low-S0/E clusters having a steeper elliptical
distribution.

\hbox{~}
\centerline{\psfig{file=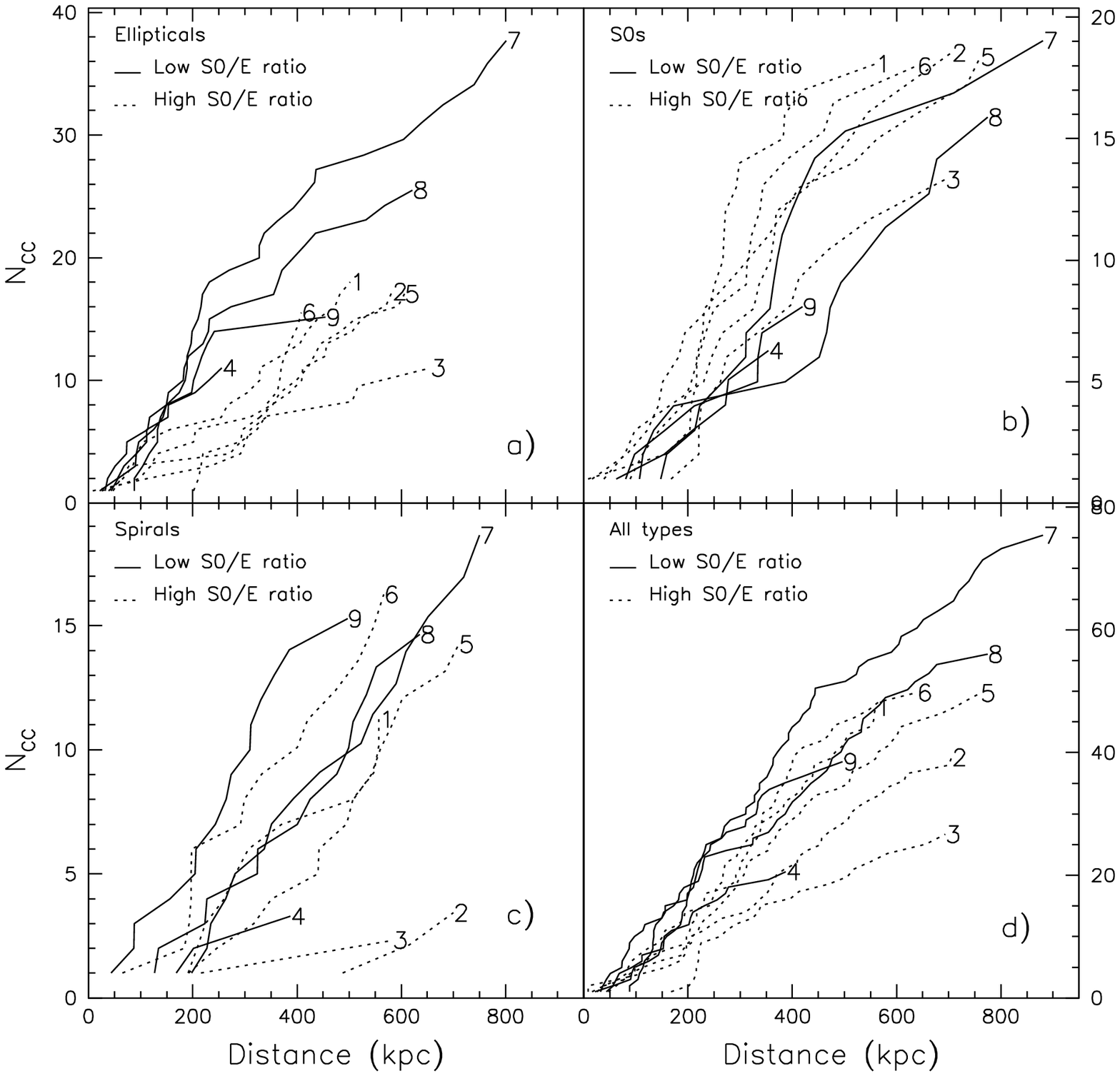,width=7.0in}}
\noindent{\scriptsize
\addtolength{\baselineskip}{-3pt} 
\hspace*{0.3cm} {\bf Fig.~7.} \ 
Cumulative but un-normalized radial distributions of individual
low-S0/E and high-S0/E clusters (full and dotted lines, respectively),
corrected for area incompleteness as explained in the text.
\addtolength{\baselineskip}{3pt}
}

In the following, the clusters with a high and low concentration of
ellipticals will be indicated by the acronyms $HEC$ and $LEC$.
It is worth noting that, since in our sample there is
a perfect correspondence between the concentration of ellipticals and S0/E
ratio, our $HEC$ and $LCE$ clusters coincide with the low-S0/E and high-S0/E 
sub-sets, respectively.

\subsection{Evolution in morphological content}

We will now investigate the evolution of the galactic morphologies 
by comparing our results with other studies at lower and higher redshift.

Before performing this comparison, we have applied 3 different statistical
corrections to our raw counts in Table~4:
(i) The correction for incompleteness due to the irregular shape 
of the area surveyed was computed according to the procedure outlined
in the previous section (see caption of Fig.7).
(ii) The systematic differences in the morphological classifications
between WJC+MORPHS and GF (see \S3), were corrected for my making the
appropriate adjustments to GF's morphological counts. Specifically,  
we have multiplied GF's S0 counts by 73/57 and, in
order to preserve the total counts, we have taken the differential counts 
from both the E and Sp populations according to the relative percentages
given in section 3 (75\% and 25\% , respectively).
(iii) The field contamination was determined from the galaxy number counts 
given by Metcalfe et al.\ (1995) and the breakdown into morphological classes 
(E/S0/Sp=18:27:56) was computed for our limiting magnitudes at $z\sim
0.2$ from the fits to the differential number counts of the Medium
Deep Survey (S97). The background assumed is listed for each cluster
in Table~5 which is available in CD-ROM form. 
The number of contaminating early-type galaxies
was found to agree with the number of probable background galaxies
redder than the elliptical sequence in the color-magnitude plots, hence
local background variations should not be dramatic towards
the clusters in this sample. As it will be clear in the following discussion,
the uncertainty introduced by the field contamination has negligible
consequences on our results.

At higher redshifts, we consider the MORPHS distant cluster sample plus
five additional clusters in the range $z=0.2-0.3$. The latter include the
3 clusters at $z\sim 0.3$ from the $HST$-based morphological study of
Couch et al. (1998), together with A2218 and A1689 (both at
$z=0.18$) for which archival HST/WFPC-2 images were available. These
images were used by WJC to morphologically classify the galaxies in
A2218 and A1689, in the same way as was done for the Couch et al. and
MORPHS studies. Details of all 5 clusters are given in Table~6; hereafter
we shall refer to them as the C98+ sample.

As for our clusters, the morphological number counts
of the MORPHS and the C98+ samples have been computed down to
$M_V=-20$ and the raw counts have been corrected for 
incompleteness due to the irregular shape of the area surveyed,
(according to the procedure outlined earlier) and for field
contamination. The latter were determined from the morphological galaxy
number counts of the Medium Deep 
Survey as in S97. The magnitude limits adopted for the C98+ clusters
are listed in Table~6. The limits for the MORPHS dataset are  
$M_{lim}^{D97}-1$\,mag, 
where $M_{lim}^{D97}$ is given in Table~1 of D97.\footnote{In fact, due 
to a transcription error in D97, $M_{lim}^{D97}$ corresponds to $M_V=-19$.}
The magnitude limits were derived adopting the transformations between the
$HST$ and standard photometric bands given by Holtzman et al. (1995).
The Cousins $I$-band
calibration of the HST A1689 image was kindly provided by I. Smail.

\hbox{~}
\vspace{-2.0cm}
\centerline{\psfig{file=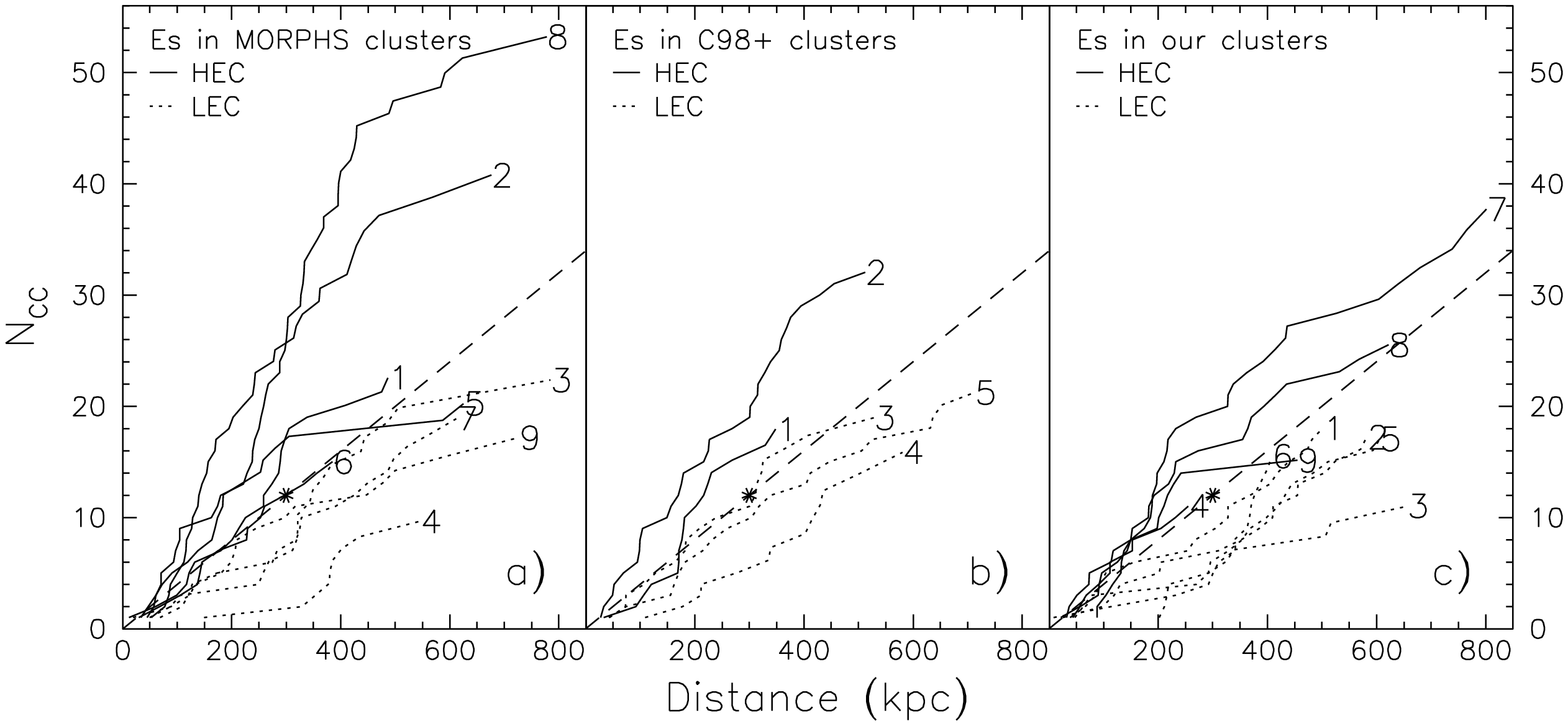,angle=-90,width=6.5in}}
\vspace{-1in}
\noindent{\scriptsize
\addtolength{\baselineskip}{-3pt} 
\hspace*{0.3cm} {\bf Fig.~8.} \
Corrected cumulative distributions of ellipticals in the MORPHS and
C98+ clusters (panel a and b, respectively), compared with
the corresponding distributions from our lower redshift sample (panel c).
Full and dotted lines represent clusters above and below the
arbitrary cut--off point $R=300$\,kpc, $N_{CC}$=12 (asterisk).
The numbers at the end of each line
identify the clusters according to the ranking order in 
Table~4 (our sample), in Table~6 (C98+ sample)
and according to the following order for the MORPHS sample:
1) Cl1447; 2) Cl0024; 3) Cl0939; 4) Cl0303; 5) 3C295; 6) Cl0412;
7) Cl1601; 8) Cl0016; 9) Cl0054. 
\addtolength{\baselineskip}{3pt}
}

To discuss the distant cluster data in the framework of the previously
noted S0/E dichotomy, we have also tried to classify the MORPHS and
the C98+ clusters according to our bimodal scheme $HEC$/$LEC$.
In Figure~8, the corrected cumulative counts ($N_{CC}$) of the elliptical
galaxies in the MORPHS and C98+ samples (Fig.~8a,b) are
compared with the distributions in our sample (Fig.~8c; see
also Fig.~7a). 
Even if no clear separation between $HEC$ and $LEC$
exists among the clusters in Figures 8a and 8b, we have tentatively
divided both samples according to the arbitrary criterion (also
working in Fig.~8c for our sample) that clusters having
$N_{CC}(R=300$\,kpc)$>$12~($<$12) belong to the $HEC$~($LEC$) family
(see asterisks and dashed lines in Fig.~8).

\begin{table*}
{\scriptsize
\begin{center}
\centerline{\sc Table 6: C98+ sample}
\vspace{0.1cm}
\begin{tabular}{lcccl}
\hline\hline
\noalign{\smallskip}
 {Cluster} & z & Area($\rm Mpc^2$) & Ref. & $M_{lim}$ \cr 
\hline
\noalign{\medskip}
1) A2218 & 0.171 & 0.4 & WJC & 19.48 F702W (HST R)\cr 
2) A1689 & 0.181 & 1.0 & WJC & 19.17 I            \cr 
3) AC118 & 0.308 & 0.6 & C98 & 20.80 F702W (HST R)\cr 
4) AC103 & 0.311 & 0.6 & C98 & 20.80 F702W (HST R)\cr 
5) AC114 & 0.312 & 1.0 & C98 & 20.80 F702W (HST R)\cr 
\noalign{\smallskip}
\noalign{\hrule}
\noalign{\smallskip}
\end{tabular}
\end{center}
}
\vspace*{-0.8cm}
\end{table*}

At low redshift, we refer to D80a and Oemler (1974; hereafter O74) as
local benchmarks.  The morphological fractions and ratios of the high- and
low--concentration nearby clusters (with a 1\,Mpc$^2$ area cut) of D80a
were obtained from Table~2 and Fig.~3 in D97.  In the following
figures we also plot the values quoted by Oemler (1974, O74) for
different cluster types.  Oemler divided clusters at low redshift into
three groups: spiral-rich (SR), elliptical-rich (ER) and S0-rich (S0R,
named spiral-poor by O74) according to their galaxy content (ER:
E:S0:Sp=3:4:2; SR: E:S0:Sp=1:2:3; S0R: E:S0:Sp=1:2:1).  We will come
back to this point in \S6; here we want to stress that Oemler's low
redshift points should be considered only as indicative because they
were not found applying the same magnitude and area limits used in
this work and have been taken from the approximate ratios given by
O74. Moreover, it is also worth stressing that the correspondence
between the D80a high(low)--concentration clusters and our
$HEC$($LEC$) clusters is far from being demonstrated.

The fully-corrected morphological fractions and ratios of all the clusters as a
function of redshift are presented in Figs.~9 and 10, highlighting the
$HEC$/$LEC$ dichotomy (filled/open symbols represent $HEC$/$LEC$
clusters respectively). 
In order to evaluate the errors associated with background subtraction,
we have computed the changes in the morphological fractions
that occur if the field correction for each cluster and galactic type
is varied by an amount equal to the correction itself (100\% error, see Table~5
available in CD-ROM form).  
The variation in the morphological fractions is on average
0.03, ranging between 0.01 and 0.08. In Figs.~9 and 10 the errorbars
represent the Poissonian errors due to the small numbers of galaxies 
and are typically greater than 0.1. Therefore the Poissonian errors 
always dominate over the errors due to field subtraction. 

In spite of the large errors, it is clear from these
figures that there are systematic trends with z: the
spiral fraction declines and the S0 fraction rises in going towards
lower redshifts. The morphological fractions in our clusters are
intermediate
between the high and the low redshift values and seem to trace a
continuous change of the abundance of S0 and spiral
galaxies. In contrast, the elliptical fraction (top panel in
Fig.~9) shows no particular trend with redshift, but rather
a large scatter from cluster to cluster at any epoch. The
mirror-like trends of the S0 and spiral fractions (Figures~9b,c) are
well represented by the behaviour of the S0/Sp ratio in
Fig.~10b. From this, we can confidently argue that, as the redshift
becomes lower, the S0 population tends to grow at the
expense of the spiral population. 
\footnote{While
this paper was being refereed, a morphological study of the cluster
CL1358+62 at z=0.33 appeared as a preprint (Fabricant, Franx \& van Dokkum
2000). The area covered in this study is bigger than the area considered
here; neverthless, the morphological fractions -- as independently
classified by the authors
and by A. Dressler -- fall in both cases within the typical values
observed at that redshift in Fig.~9.}

Figure~10a suggests the existence of a large intrinsic scatter in the
S0/E ratio, which, at least within our sample, can be mostly ascribed
to the different cluster morphologies: clusters showing a marked
concentration of ellipticals in the central region have lower S0/E
ratios with respect to the clusters in which E galaxies are more or
less uniformly distributed inside the cluster area. A similar tendency
is possibly seen at higher redshift as well and at low redshift
comparing elliptical-rich and S0-rich clusters. The dotted and dashed
straight lines in Figure~10a represent the formal least--square weighted
fits (0.09$<$z$<$0.65) to the HEC and LEC clusters, respectively. In
spite of the scatter, likely due to the intrinsic dependency of the
S0/E ratio on the cluster morphology, the trend with redshift found by
D97 (full line in Figure~10a) turns out to be confirmed.

\hbox{~}
\vspace{-2.0cm}
\centerline{\psfig{file=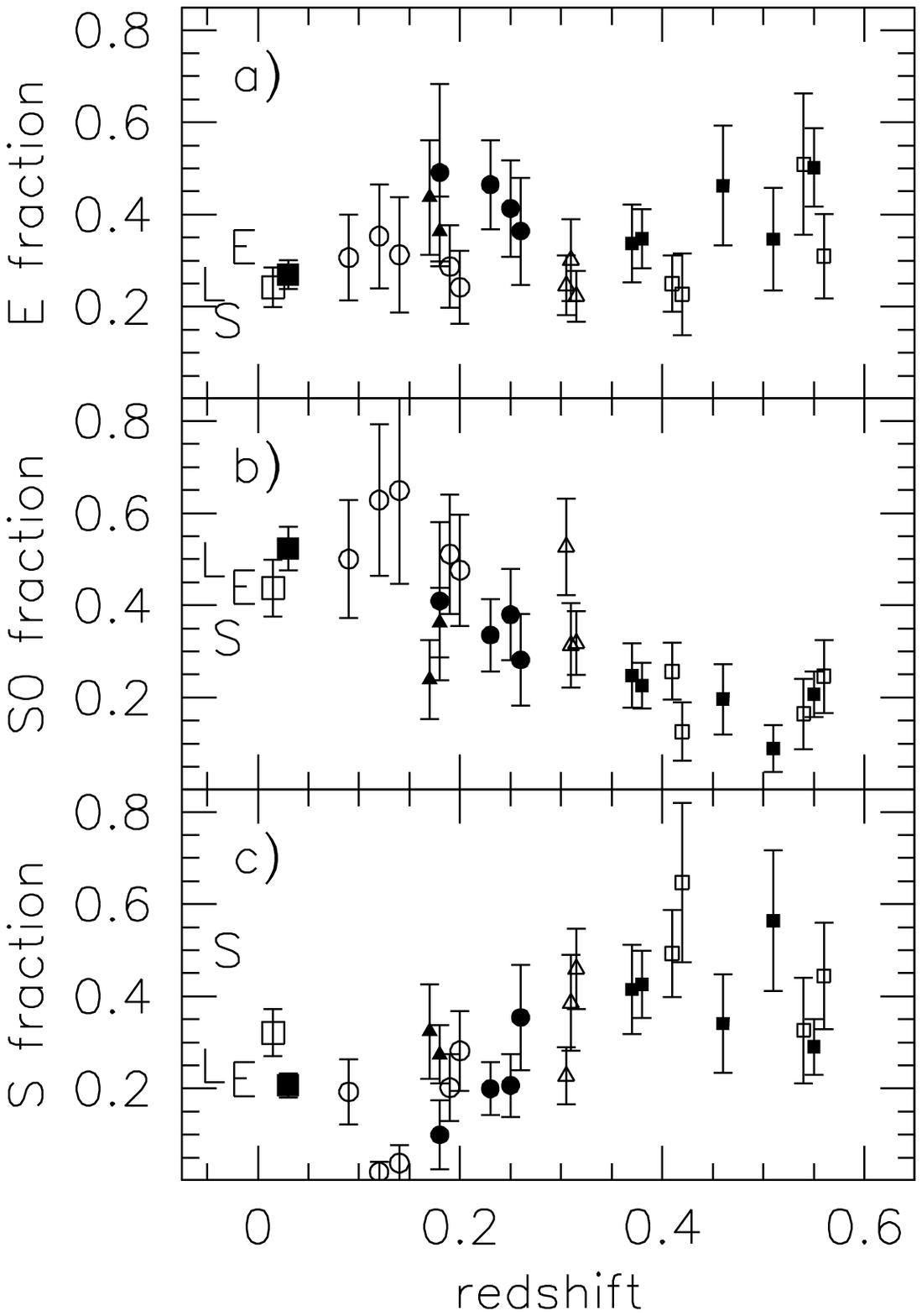,angle=0,width=6.5in}}
\vspace{-0.2in}
\noindent{\scriptsize
\addtolength{\baselineskip}{-3pt}
\hspace*{0.3cm} {\bf Fig.~9.} \ 
Morphological fractions as a function of redshift. $HEC$ and $LEC$ clusters
are displayed as solid and open symbols, respectively. The
values from our sample are indicated by circles, whereas those
from the MORPHS and C98+ samples are indicated with squares
and triangles, respectively. All these data are corrected both for field 
contamination and for the irregular shape of the area surveyed (see text).
The errorbars correspond to Poissonian values.
The average values derived for high- ({\it large solid squares})
and low--concentration ({\it large open squares}) clusters from D80 
(see text) are plotted at $z \sim 0$. 
Oemler's datapoints are indicated with letters (E = elliptical-rich;
L = S0-rich; S = spiral-rich) and are placed at $z<0$ for display purposes. 
\addtolength{\baselineskip}{3pt}
}

\hbox{~}
\vspace{0.8in}
\centerline{\psfig{file=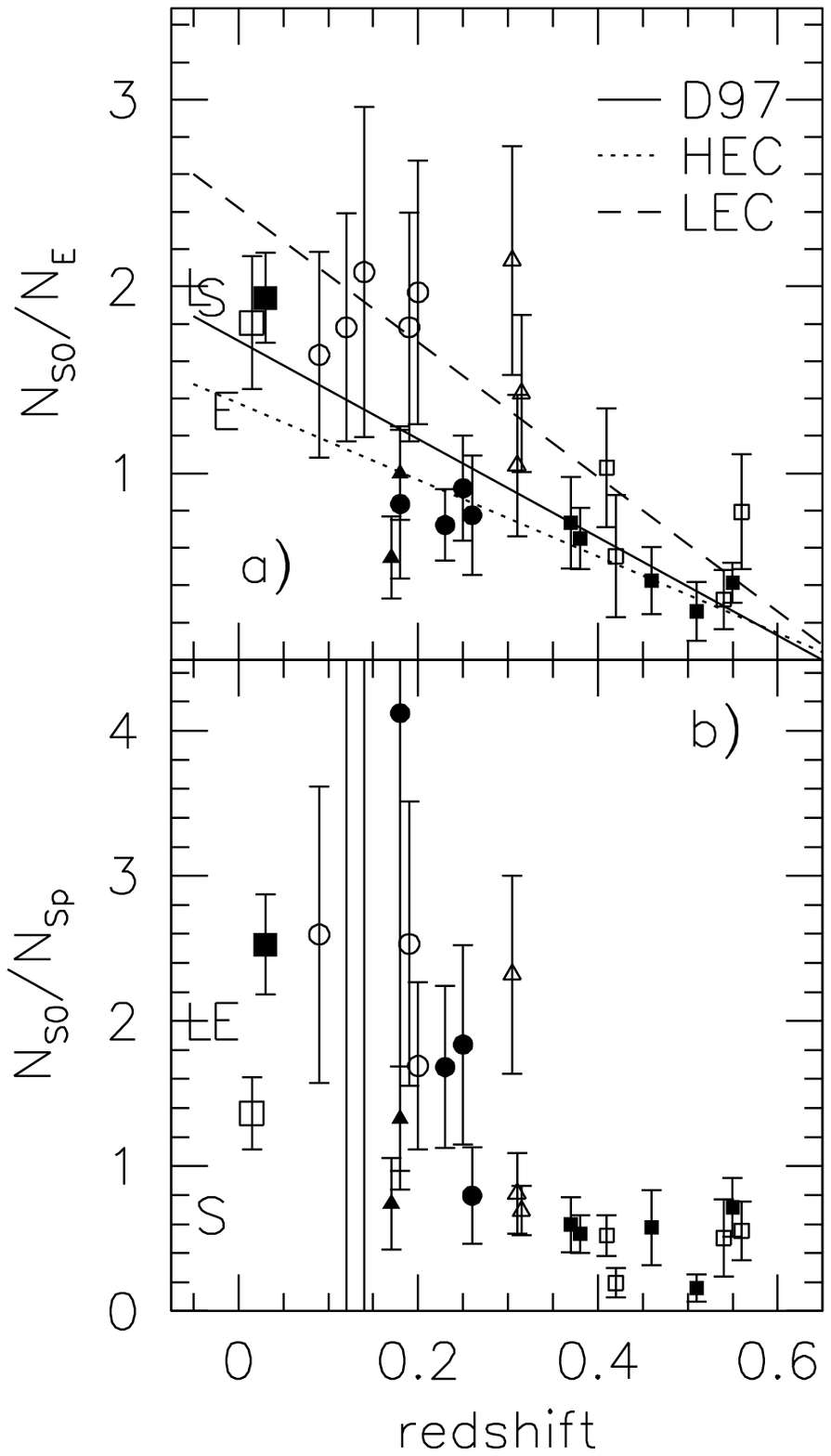,angle=0,width=6.5in}}
\vspace{-0.2in}
\noindent{\scriptsize
\addtolength{\baselineskip}{-3pt} 
\hspace*{0.3cm} {\bf Fig.~10.} \ 
The S0/E and S0/Sp ratios as a function of redshift.  The meaning of
the symbols is as in Fig.~9.  The dotted and dashed lines in the top
panel represent the least square weighted fits (0.09$<$z$<$0.65) to
the HEC and LEC clusters, respectively. The linear regression of the
MORPHS data given by D97 is represented by the {\it solid} line.
\addtolength{\baselineskip}{3pt}
}


\section{The morphology-density relation}

In this section we examine the relative occurrence of each
morphological type as a function of the local projected density of
galaxies (i.e. the morphology-density relation). The local densities
have been computed -- following D80a and D97 -- in a rectangular area
containing the 10 nearest neighbors.  In Figure~11 we present the
local density distributions of galaxies of the different morphological
types (upper panels), together with the corresponding distributions of
morphological fractions (lower panels) for our global cluster sample
(leftmost panels), as well as the 
$LEC$ and $HEC$ sub-samples (middle and rightmost panels).

\hbox{~}
\centerline{\psfig{file=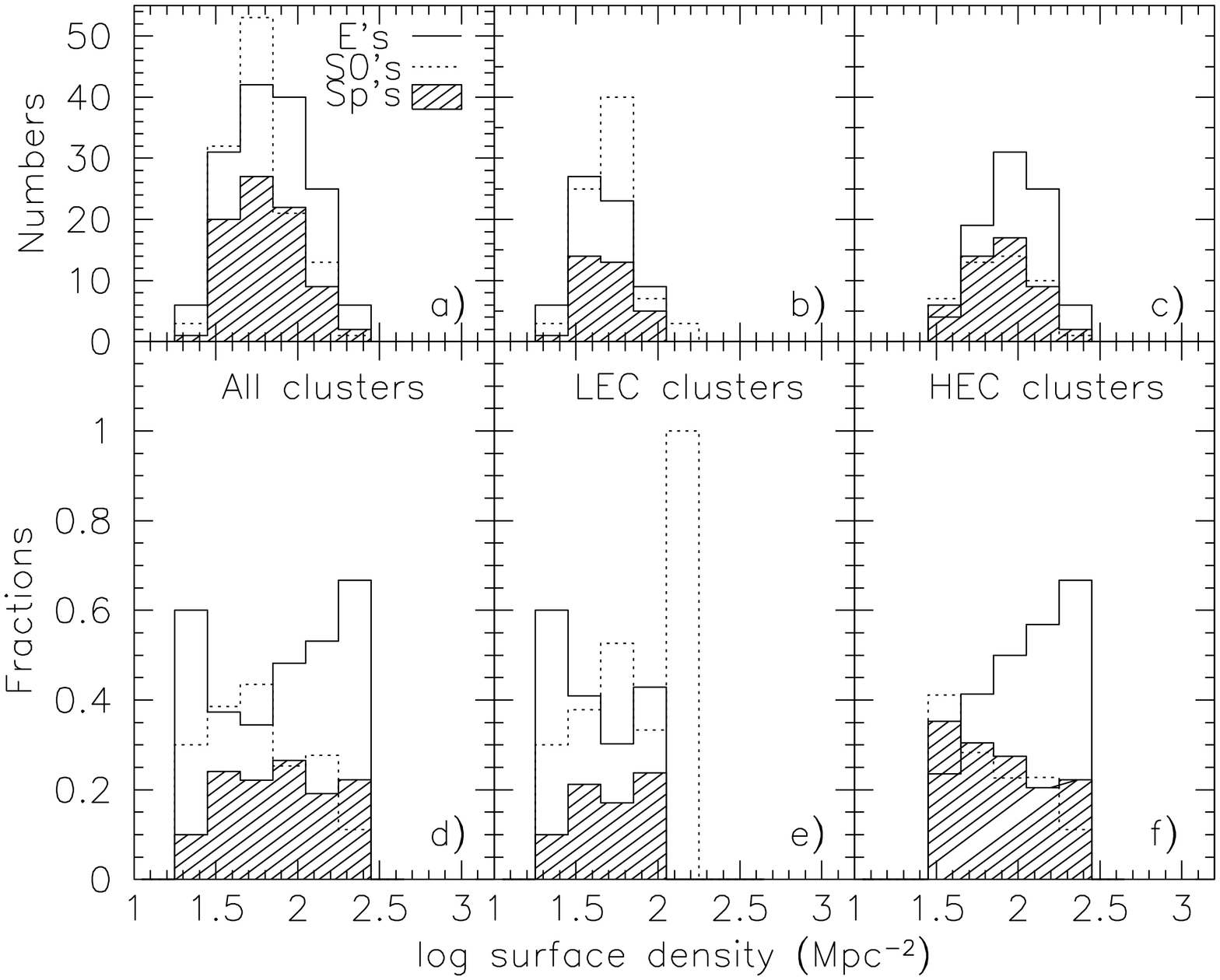,angle=-90,width=6.5in}}
\noindent{\scriptsize
\addtolength{\baselineskip}{-3pt} 
\hspace*{0.3cm} {\bf Fig.~11.} \ 
Number and fraction of galaxies of different morphological types
as a function of the local density for the 9 clusters in our sample
(leftmost panels) and for the sub-samples of $LEC$ (middle panels)
and $HEC$ (rightmost panels) clusters. The distributions of
elliptical, S0 and spiral galaxies are indicated with a full line,
a dotted line and a shaded histogram, respectively. 
\addtolength{\baselineskip}{3pt}
}

Apart from the different value of the global S0/Sp ratios, our
distributions (Fig.~11d, e and f) look qualitatively very similar to
those of the MORPHS sample (Figs.~4, ~5 and ~6 in D97), particularly
if we compare our $HEC$ and $LEC$ cluster families with the 
high- and low-concentration families in D97.  As in the higher
redshift clusters, a morphology-density relation is present in highly
concentrated clusters and absent in the low concentration clusters
where a possible small {\it anticorrelation} can be noted both here
and in D97. The fact that the density distributions at $z \sim 0.1$
resemble those at $z \sim 0.5$ much more closely than those at $z \sim
0$ could suggest that the morphology-density relation in low
concentration clusters was established in the last 1-2 Gyr, but a definite
conclusion cannot be reached on the basis of the available data.

Another remarkable feature of the MD distributions is that, moving
from high redshift to the intermediate redshift regime, the
fraction of Es as a function of the local density appears to be
practically unchanged for both $HEC$ and $LEC$ clusters, i.e. the
elliptical-density relations at $z=0.2$ are also {\it quantitatively}
similar to that of the MORPHS sample at $z=0.5$. We also note that the
$LEC$ distributions remain relatively flat over this redshift interval, 
while a strong increase in the S0/Sp fraction takes
place (compare our Fig.~12e with Fig.~8 in D97). This suggests that in
the $LEC$ clusters the Sp$\rightarrow$S0 transformation process is
highly efficient and that the efficiency is largely independent of
local density. In contrast, the S0-- and spiral--density distributions in
$HEC$ clusters at the two different redshifts
(compare our Fig.~11f with Fig.~6 in D97), have quite different 
slopes, suggesting that the process of
transfer from spirals to S0s is more efficient in low density regions
than in high density ones. The above analysis is based on the
comparison of two rather small cluster samples and, therefore, it
needs to be confirmed by the morphological study of more sizeable
samples. Nevertheless, there is the suggestion that, at least in our
range of observed local densities (1.3$\ls\log\rho\ls$2.4), the efficiency
of the possible transition from spiral to S0 morphology seems to increase
with decreasing local density.

Finally, the fact that no MD relation is found in our $LEC$ clusters
demonstrates that the dichotomy in the S0/E ratio discussed in the
previous section cannot merely be ascribed to a universal MD relation
at $z=0.2$ combined with different density ranges for the $HEC$ and
$LEC$ clusters: the two types of clusters are intrinsically different
both in their global morphological content and in the arrangement of
the morphological types as a function of the local density.

\section{Conclusions}

1) The morphological properties of the galaxy populations in nine
clusters at $z=0.1-0.25$ are found to be intermediate between those at
$z\sim 0.4-0.5$ and those at low-z, with a moderate spiral content and
a moderate ``deficiency'' (as compared to lower redshifts) of S0
galaxies. Our results support the evolutionary scenario, inferred
from higher redshift studies (D97, S97, Couch et al. 1998), involving
the disk galaxy populations in which there is a progressive
morphological conversion in clusters, from spirals into S0's.

2) At $z\sim 0.2$, we find a dichotomy in the relative occurrence of S0
and elliptical galaxies: four of our clusters display a low S0/E ratio
($\sim 0.8$) while two of our clusters have a significantly higher
ratio ($\sim 1.9$) similar to the other clusters in our sample at
$z=0.1$.  The most likely interpretation of this dichotomy and of the
large scatter in the S0/E ratio at $z\sim 0.2$ is that such a ratio is
both a function of the redshift and of the cluster ``type'', being
significantly lower in clusters with a strong central concentration of
elliptical galaxies.

3) At $z\sim 0.1-0.2$ a morphology-density relation exists only for the
high-concentration clusters and is absent in the low-concentration
ones. The same result was found at $z \sim 0.5$ (D97), while at low
redshift the correlation between galaxy morphology and local density
is present in all types of clusters of the D80a sample. Although the number of
galaxies is too small to draw definite conclusions, these results seem
to suggest that the morphology-density relation in low-concentration
clusters was established only in the last 1-2 Gyr, but only 
additional data and a homogeneous systematic analysis
both at low and moderate redshifts will clarify this matter. Moreover,
comparing our MD relations with the corresponding ones at $z \sim 0.5$
(D97), we suggest that the efficiency of the Sp$\rightarrow$S0
transformation process anticorrelates with the local density.

The relation between the S0/E ratio and the spatial concentration of the 
ellipticals is not surprising in the light of the well-known
correlations between the galaxy content and the cluster type
in low-redshift clusters. As mentioned in \S4, Oemler (1974) 
grouped clusters in three classes: spiral-rich (SR), elliptical-rich
(ER, the most spherical in shape and concentrated) and S0-rich (S0R,
named ``spiral-poor'' by O74, ``not quite as centrally concentrated as
the ER class, but more regular than the SR class'').
Interestingly, O74 suggested that S0-rich clusters are dynamically
evolved clusters (they already show segregation by mass and
morphological type) representing a later evolutionary stage of
spiral-rich clusters, following the evolution of a significant
fraction of the spiral galaxies into S0's. In contrast, in the
scenario proposed by O74, E-rich clusters are well evolved but
\sl intrinsically different \rm from the S0R clusters: 
although possibly the dynamically oldest
type of clusters, their high elliptical content implies that
they did not evolve from the spiral-rich clusters and is likely due to 
an enhanced formation rate of ellipticals in regions that began
as the densest fluctuations in the early universe.

We speculate that in our sample at $z \sim 0.2$, the four clusters with 
a strong central concentration of ellipticals (and the lowest S0/E ratios)
are presumably the analogues (and progenitors) of the low-z E-rich clusters
(see the extrapolation at low redshift of dotted line in Figure~10a),
while the low-concentration clusters (with the highest S0/E ratio)
seem to be the analogues of the ``S0-rich'' clusters. 

The effects of redshift (evolution) and cluster type are expected
to mingle in various proportions at the different epochs.
Following O74, we suggest that the relative occurrence of S0's and spirals
is mostly linked with the ``maturity'' of the cluster with spirals 
progressively evolving into S0's,
while the ellipticals are well in place at redshifts greater than those 
considered here and their abundance and concentration reflect an
``original imprinting'' (see also D97). Then, the S0/Sp ratio should
be related to the evolutionary epoch of the cluster and the S0/E value
should be determined both by the epoch and the cluster type
(nurture and nature, in a way),
with the redshift being the dominant effect at early epochs.
\footnote{If the ER and S0R clusters are the endpoints of the evolution of two
originally-different types of clusters and both types experienced the
accretion of large numbers of spirals that with time turned into S0's,
then when looking further back in time, the difference in the S0/E
ratio between the precursors of the ER and of the S0R clusters should
become smaller and smaller: at $z\sim 0.4-0.5$ all types of clusters
are expected to display a low S0/E ratio (with smaller fluctuations
among the different types of clusters), as indeed is observed in the
MORPHS dataset, simply because many of the S0's have not yet formed.}

\section*{Acknowledgements} 

Based on observations made with the Nordic Optical Telescope, La
Palma, and the Danish 1.5-m telescope at ESO, La Silla, Chile.  The
Nordic Optical Telescope is operated jointly by Denmark, Finland,
Iceland, Norway, and Sweden, in the Spanish Observatorio del Roque de
los Muchachos of the Instituto de Astrofisica de Canarias.

The authors are grateful to the anonymous referee for 
the prompt refereeing that helped us to improve the rigor
and clarity of this paper.
BMP and WJC warmly thank Ian Smail for providing the images and
the photometric catalog of A1689, for his valuable assistance in
this project, for carefully reading the manuscript
and suggesting a number of changes that improved the paper. They are also
thankful to their collegues of the MORPHS group for many interesting 
discussions and useful advice.

This work was supported by the Formation and Evolution of Galaxies
network set up by the European Commission under contract ERB
FMRX-CT96-086 of its TMR program and by the Danish Natural Science
Research Council through its Centre for Ground-Based Observational
Astronomy. WJC acknowledges the Schools of Physics at Bristol
and St Andrews Universities and the European Southern Observatory
for their hospitality during the course of this work.
This research has made use of the NASA/IPAC Extragalactic Database
(NED) which is operated by the Jet Propulsion Laboratory, Caltech,
under contract with the National Aeronautics and Space Administration.

\smallskip


\newpage
\begin{thebibliography}{}
\itemsep=0in

\bibitem{}
Abadi, M.\,G., Moore, B., Bower, R.\,G., 1999, MNRAS, 308, 947

\bibitem{}
Andreon, S., Davoust, E., Heim, T., 1997, A\&A, 323, 337

\bibitem{}
Andreon, S., 1998, ApJ, 501, 533

\bibitem{}
Barger, A.\,J., Arag\`on-Salamanca, A., Smail, I.,
Ellis, R.\,S., Couch, W.\,J., Dressler, A., Oemler, A.\ Jr, Poggianti, B.\,M, 
Sharples, R.\,M., 1998, ApJ, 501, 522 

\bibitem{}
Bautz, L.P., Morgan, W.W, 1970, ApJ, 162, L149

\bibitem{}
Bertin, A., Arnouts, S., 1996, A\&AS, 117,393

\bibitem{}
Butcher, H., Oemler, A.\ Jr. 1978, ApJ, 226, 559

\bibitem{}
Butcher, H., Oemler, A.\ Jr. 1984, ApJ, 285, 426

\bibitem{}
Capaccioli, M., Vietri, M., Held, E.V., Lorenz, H., 1991, ApJ, 371, 535

\bibitem{}
Couch, W.\,J., Barger, A.\,J., Smail, I., Ellis, R.\,S., Sharples, R.\,M., 
1998, ApJ, 497, 188

\bibitem{}
Couch, W.\,J., Ellis, R.\,S., Sharples, R.\,M., Smail, I., 1994, ApJ, 430,
 121

\bibitem{}
Dressler, A., 1980a, ApJ, 236, 351 (D80)

\bibitem{}
Dressler, A., 1980b, ApJS, 424, 565 (DCAT80)

\bibitem{}
Dressler, A., Oemler, A.\ Jr., Butcher, H., Gunn, J.\,E., 1994, ApJ, 430, 107

\bibitem{}
Dressler, A., Oemler, A.\ Jr., Couch, W.\,J., Smail, I.,
Ellis, R.\,S., Barger, A., Butcher, H., Poggianti, B.\,M., Sharples,
R.\,M., 1997, ApJ, 490, 577 (D97)

\bibitem{}
Dressler, A., Smail, I., Poggianti, B.\,M., Butcher, H., Couch, W.\,J.,
Ellis, R.\,S., Oemler, A.\ Jr., 1999, ApJS, 122, 51 

\bibitem{}
Ebeling, H., Edge, A., Bohringer, H., Allen, S., Crawford, C.,
Fabian, A., Voges, W., Huchra, J., 1998, MNRAS, 301, 881

\bibitem{}
Ellis, R.\,S., Smail, I., Dressler, A., Couch. W.\,J., Oemler, A.\ Jr.,
Butcher, H., Sharples, R.\,M., 1997, ApJ, 483, 582

\bibitem{}
Fabricant, D., Franx, M., van Dokkum, P., 2000, ApJ, in press 
(astro-ph 0003360)

\bibitem{}
Fasano, G., Bettoni, D., D'Onofrio, M., Kj\ae rgaard, P., Moles, M., 2000, 
in preparation

\bibitem{}
Fasano, G., Vio, R., 1991, MNRAS, 249, 629 

\bibitem{}
Franceschini, A., Silva, L., Fasano, G., Granato, G.L., Bressan, A., Arnouts,
S., Danese, L., 1998, ApJ, 506, 600 

\bibitem{}
Frei, Z., Guhathakurta, P., Gunn, J.\,E., Tyson, J.\,A., 1996, AnJ,
111, 174

\bibitem{}
Holtzman, J.\,A., Burrows, C.\,J., Casertano, S., Hester, J.\,J., Trauger, J.\,T., Watson, A.\,M., Worthey, G., 1995, PASP, 107, 1065

\bibitem{}
Jones, C., Forman, W., 1999, ApJ, 511, 65

\bibitem{}
Jorgensen, I., 1994, PASP, 106, 967

\bibitem{}
Kelson, D.\,D., van Dokkum, P.\,G., Franx, M., Illingworth, G.\,D.,
Fabricant, D., 1997, ApJ, 478, L13

\bibitem{}
Kelson, D.\,D., Illingworth, G.\,D., van Dokkum, P.\,G., Franx, M.,
1999, ApJ, in press (astro-ph 9906152)

\bibitem{}
Landolt, A.U., 1992, AJ, 104,340

\bibitem{}
Lavery, R.\,J., Henry, J.\,P., 1988, ApJ, 330, 596

\bibitem{}
Lavery, R.\,J., Pierce, M.\,J., McClure, R.\,D., 1992, AJ, 104, 2067

\bibitem{}
Lavery, R.\,J., Henry, J.\,P., 1994, ApJ, 426, 524

\bibitem{}
Lubin, L.\,M., Postman, M., Oke, J.\,B., Ratnatunga, K.U., Gunn, J.\,E.,
Hoessel, J.\,G., Schneider, D.\,P., 1998, ApJ, 116, 584

\bibitem{}
Mathis, J.\,S., 1990, ARAA, 28, 37

\bibitem{}
Metcalfe, N., Shanks, T., Fong, R., Roche, N., 1995, MNRAS, 273, 257

\bibitem{}
Moore, B., Katz, N., Lake, G., Dressler, A.,  Oemler, A.\ Jr. 1996,
Nature, 379, 613

\bibitem{}
Moore, B., Lake, G., Katz, N., 1998, ApJ, 495, 139

\bibitem{}
Oemler, A.\,Jr., 1974, ApJ, 194, 1

\bibitem{}
Oemler, A.\ Jr., Dressler, A.,  Butcher, H., 1997, ApJ, 474, 561

\bibitem{}
Poggianti, B.\,M., 1997, A\&AS, 122, 399

\bibitem{}
Poggianti, B.\,M., Smail, I., Dressler, A., Couch, W.\,J., Barger, A.\,J.,
Butcher, H., Ellis, R.\,S., Oemler, A.\,Jr., 1999, ApJ, 518, 576 

\bibitem{}
Pignatelli, E.,Fasano, G., 1999, Proceedings: {\it First Italian Workshop 
of the "Network sulla formazione ed evoluzione delle galassie"},
http://www.brera.mi.astro.it/docB/galaxy/news.html

\bibitem{}
Rood, H.J., Sastry, G.N., 1971, PASP, 83, 313

\bibitem{}
Simien, F., de Vaucouleurs, G., 1986, ApJ, 302, 564

\bibitem{}
Smail, I., Dressler, A., Couch, W.\,J., Ellis, R.\,S., Oemler,
A.\ Jr, Butcher, H., Sharples, R.\,M., 1997, ApJS, 110, 213 (S97)

\bibitem{}
Smail, I., Edge, A.\,C., Ellis, R.\,E., Blandford, R.\,D., 1998, MNRAS,
293 124

\bibitem{}
Thompson, L.\,A., 1986, ApJ, 300, 639

\bibitem{}
Thompson, L.\,A., 1988, ApJ, 324, 112

\bibitem{}
van Dokkum, P.\,G., Franx, M., 1996, MNRAS, 281, 985

\bibitem{}
van Dokkum, P.\,G., Franx, M., Kelson, D.\,D., Illingworth, G.\,D., Fisher,
D., Fabricant, D., 1998, ApJ, 500, 714

\bibitem{}
van Dokkum, P.G., Franx, M., Fabricant, D., Kelson, D.\,D., 
Illingworth, G.\,D., 1999, ApJ, 520, L95

\bibitem{}
van Dokkum, P.G., Franx, M., Fabricant, D.,
Illingworth, G.\,D., Kelson, D.\,D., 2000, ApJ, in press
(astro-ph 0002507)

\bibitem{}
Wirth, G.\,D., Koo, D.\,C., Kron, R.\,G., 1994, ApJ, 435, L105

\end{thebibliography}
\end{document}